# Principles and Overview of Network Steganography

Józef Lubacz, Wojciech Mazurczyk, Krzysztof Szczypiorski
Institute of Telecommunications, Warsaw University of Technology, Warsaw, Poland
e-mail: {jl, wm, ksz}@tele.pw.edu.pl

**Abstract.** The paper presents basic principles of network steganography, which is a comparatively new research subject in the area of information hiding, followed by a concise overview and classification of network steganographic methods and techniques.

## 1. Introduction

As the production, storage and exchange of information becomes more extensive and important in the functioning of societies, the problem of protecting the information from unintended and undesired use becomes more complex. In modern societies, protection of information involves many interdependent policy and technological issues relating to information confidentiality, integrity, anonymity, authenticity, utility, etc. Classification of these issues is a problem in itself. Several such classifications have been proposed; among the most popular classifications are the CIA Triad (Confidentiality, Integrity and Availability) and the Parkerian Hexad [9]. In this paper, we do not discuss this type of conceptualisations; we focus on methods for providing confidentiality in the communication of digitised information. Our goal is to characterise a subset of such methods named *network steganography* which are information hiding techniques that utilise network protocols as enablers of hidden communication. The term network steganography was first coined in [12]. We begin our considerations with some general remarks concerning the terminology used in the area of information security, as we believe there is some confusion in this area. In particular, we aim to clarify the relationships between the terms steganography, cryptography and information hiding.

The terms steganography and cryptography originate from the ancient Greek words *steganos*, meaning protected (covered), and *kryptos*, meaning hidden (secret), respectively. These two meanings are obviously quite close; if considered literally, the term steganography could be substituted with the term cryptography, and vice versa. It is also plausible to consider both steganography and cryptography to be different methods of hiding information: steganographic methods hide information, thereby making it "difficult to notice" (by means of embedding it in an information carrier), while cryptographic methods hide information by making it "difficult to recognize" (by means of transforming it). Note also that messages are bearers of information that is communicated (or stored); thus, the terms "information hiding" and "message hiding" (as well as the terms "communication" and "transmission") should be formally distinguished. Herein, these distinctions are made only in the case of possible confusion.

Such considerations of terminology, although potentially helpful in distinguishing different methods for providing information security, meet obstacles due to established conventions. In particular, in 1996, at the first Information Hiding Workshop held in Cambridge, UK, the meaning of the term information hiding and an associated classification of techniques were agreed upon [11]. This classification encompasses steganography, but not cryptography. Apart from steganography, "anonymity", "copyright marking" and "covert channels" were distinguished. This seems a semantically inconsistent and misleading classification: (1) providing for anonymity of communicating parties can be achieved by means of hiding information that identifies the communicating parties, but also, e.g., by clever routing of data transfer; (2) copyright marking, encompassing watermarking, does not always require hiding the mark, as in some applications, it is important to have the mark clearly visible (but hard to remove); (3) the distinction (as in [10], [15]) between steganography (and network steganography in particular) and "covert channels" (supposedly defined after Lampson [5] as communication paths that were neither designed nor intended for information transfer) is not well grounded. It is plausible to say that network steganography techniques create channels for hidden communication, but not that such channels exist a priori in a communication network. The capacity of the created channels depends on the steganography technique applied in the context of a particular communication network.

Steganographic techniques may be classified with respect to different criteria, in particular with respect to their intended use and to the type of information carrier utilised. We constrain our considerations to techniques based on various functions of communication protocols of contemporary communication networks. This specific class of techniques is referred to as network steganography. Steganographic techniques for storing messages may be formally considered to be network steganography methods because the stored information may eventually be transmitted. While such techniques are quite extensively discussed in the literature, they are of marginal interest in the context of our considerations.

Both cryptography and steganography techniques are practically applied in imperfect communication environments imposed by physical features of information carriers. While this imperfectness is generally an obstacle for cryptography, it is an essential enabling condition for many network steganography techniques that utilise redundant communication mechanisms (protocols) to cope with such imperfect environments to provide reliable communication.

In principle, a message to be hidden with the use of a steganographic technique may be first encrypted with some cryptographic technique. Note, however, that if applied, this will potentially increase the probability that the message is noticed and thus reduces the chance of achieving the principal goal of the steganographic method in use. The main objectives and potential applications of information hiding with the use of steganographic and cryptographic methods should not be regarded as competitive and/or complementary alternatives. This is an important point in understanding the potential benefits and threats of using steganographic and cryptographic techniques.

In contrast to cryptographic methods, the goal of steganography methods is to make the secret communication hard to notice, and this, in particular, is the reason for its alleged use by terrorists and by citizens of countries that prohibit the use of cryptographic methods in communication. Thus, research on steganography, and on network steganographic techniques in particular, provokes ethical questions concerning the risk that the research results will be used for malicious purposes. There is (should be) an ethical obligation for researchers who propose and publish new steganographic techniques to present methods for uncovering their use, i.e. steganalysis methods. As explained in the following, there is a trade-off between the effectiveness of any network steganographic technique (in terms of potential steganographic capacity) and its susceptibility to steganalysis (i.e., to being uncovered). Essentially, evaluating the effectiveness of a proposed technique requires evaluating its robustness to steganalysis. Thus, in an ethically proper presentation of a new technique, fulfilment of the ethical obligation comes naturally.

## 2. Network steganography basics

Generally speaking, when considering any communication network, three basic functionalities may be distinguished: services/applications, transport of information and information flow control. In the traditional PSTN/ISDN, i.e., circuit-switched networks, the services/applications are basically provided by the network, transport takes place through transparent channels, and the control and transport functions are virtually separated. Once the end-to-end connection and transport channel are established, information (voice or data) is transported through the network without interference. The network user has little influence on the service delivered by the network and no influence on the flow of information. The Internet, i.e., a packet switched network, has substantially changed the traditional circuit-switched network paradigm: services/applications are created by the network users rather than by the network itself, and the transport and control functions are not separated and can be influenced by the user. This change of paradigm was one of the main sources of the tremendous success of the Internet. However, these advances also introduced well-known problems with quality of service and with protecting the network and its users from harmful/undesired interference. It is thus not surprising that the Internet opened many new options for covert communication. This observation may be generalised to practically all types of contemporary fixed and mobile networks, and particularly to communication protocols, which are becoming increasingly diverse and complex, and thus susceptible to manipulation. Network steganography techniques take advantage of this susceptibility.

Even elementary functions of communication protocols can be utilised to construct a steganographic method. Consider, for example, a query-response type of exchange of messages for which the communication protocol assumes that the response should come within a specific time limit; otherwise, it is treated as excessively delayed and discarded. Communicating parties that want to use this protocol for steganographic purposes may make an agreement, which becomes their shared secret, that the responses carrying hidden information will be purposefully excessively delayed and that such responses will be read by the recipient (i.e., not discarded). Therefore, the communication protocol is manipulated for steganographic purposes. This "trick" may be effective only if the communication channel introduces some delay in the message transmission. Potential observers of the communication – potential attackers – who know the communication protocol and follow it during observation, do not become suspicious of the existence of hidden communication if the frequency of occurrence of excessively delayed responses is not considered to be abnormal, i.e., does not exceed some expected frequency that the observers assume, based on their knowledge of delay properties of the communication network.

This simple example represents some basic features of network steganography techniques. If generalised, the following features of network steganography techniques may be formulated:

**(C1)** some functions of communication protocols are modified;

**(C2)** the modification pertains to:
  **(C2a)** functions of the protocols that are introduced to cope with the intrinsic imperfectness of communication channels (errors, delays, etc.)
and/or to
  **(C2b)** functions of the protocols that are introduced to define the type of information exchange (e.g. query-response, file transfer, etc.) and/or to adapt the form of messages (e.g. fragmentation, segmentation, etc.) to the information transmission carrier;

**(C3)** the modifications are utilised by the communicating parties to make the observable effects of modifications difficult to discover (e.g., to seem to result from the imperfectness of the communication network and/or protocols).

Conditions C1, C2 and C3 constitute a proposed definition of network steganography techniques.

Note that if condition C1 is not fulfilled, i.e., if there is no interference in the communication protocol, then some form of hidden communication still may be performed, namely if the secret shared by the sender and receiver is of the form: messages a, b, c, …, are interpreted as x, y, z …. Such hidden communication cannot be discovered by observing the exchange of messages, as these are interpreted on the semantic/pragmatic level by the sender and receiver. In effect, such hidden communication can be discovered only if the shared secret is disclosed. Obviously, this is not a very interesting case for research.

Condition C2 refers to the fact that real world communication protocols must realise functions (C2a) that provide required quality-related performance of communication and functions (C2b) that govern the "logic" of communication and adapt the messages to the format of transmission carriers. If the communication functions are decomposed into functional layers, as for example in the OSI RM (Open Systems Interconnection Reference Model), then C2a functions are associated with lower layers and C2b functions with upper layers. In Fig. 1, these functions, in association with OSI RM protocol layers, are characterised in a general manner.

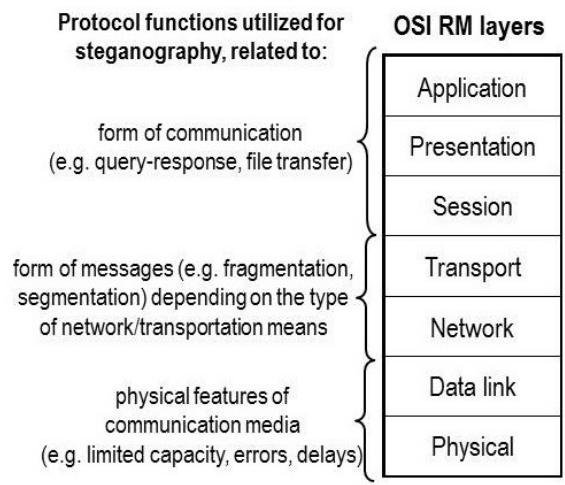

**Fig. 1** Protocol functions used for network steganography, associated with OSI RM layers

The effectiveness of a particular technique depends on how successfully the C3 condition is fulfilled. Two essential measures of the effectiveness can be considered: the potential throughput of hidden messages – referred to as steganographic bandwidth – and the resistance to discovering the hidden communication, i.e., resistance to steganalysis. The two measures are interdependent: usually, the higher steganographic bandwidth, the lower the resistance to steganalysis. The latter is difficult to estimate quantitatively and depends not only on the sophistication of a steganographic technique but also on the knowledge and efficacy of potential observers of the communication.

### 3. An overview of network steganography techniques

As this is a short paper, the following overview does not cover all of the techniques proposed in the literature. The selection of techniques discussed herein is somewhat arbitrary;

nevertheless, the authors tried to select techniques that seem to be most representative in their class or that have introduced novel ideas and inspired new directions of research.

Classification of network steganography methods may be based on protocol functions associated with the OSI RM layers (Fig. 2). In the physical and data link layers, steganographic techniques can exploit physical features of communication channels and/or imitate their imperfectness. For example, Szczypiorski and Mazurczyk [13] proposed a physical layer method called WiPad (Wireless Padding) intended for IEEE 802.11 OFDM networks. Its functioning is based on the insertion of hidden data into the padding of transmitted symbols. Szczypiorski [12] introduced a data link layer method called HICCUPS (Hidden Communication System for Corrupted Networks). The main idea of the method is to utilise transmission frames with intentionally wrong checksums. In a WLAN, all user terminals can "hear" data contained in frames transmitted in the medium; generally, frames with wrong checksums are discarded by terminals, whereas terminals that are aware of the use of the steganographic method can read such frames and extract hidden data from the frame payload field.

Network steganography methods can also use the adjustment of the form of the messages to the type of network or means of transport. Kundur and Ahsan [4] proposed two such approaches for the OSI RM layer. In one solution, the bits of hidden data are inserted into unused or reserved parts of packet headers. This method works because many protocol standards do not mandate specific/standard values for the unused and reserved parts (these are not verified at the receiver). In particular, Kundur and Ahsan proposed steganographic use of the IP header's DF (Don't Fragment) flag. This approach is successful if the sender sends packets of size that are smaller than the path's MTU (Maximum Transfer Unit). Many similar methods were also introduced for other fields of various protocols; e.g., Handel and Sandford [2] proposed using the unused bits of the IP header's ToS (Type of Service) field and of the TCP header's Flags field. In another solution, Kundur and Ahsan proposed utilising per packet sequence numbers of IP packets and their intentional re-sorting.

For the OSI RM transport layer, Mazurczyk et al. [7] introduced the RSTEG (Retransmission Steganography) technique intended for a broad class of protocols with retransmission mechanisms. The main idea of RSTEG is to not acknowledge successfully received TCP segments to intentionally invoke retransmission. The retransmitted segment of user data carries hidden data in the payload field.

Mazurczyk and Szczypiorski [8] suggested the utilisation of unused fields of the session layer protocol SIP (Session Initiation Protocol). For the presentation layer, Bender et al. [1] introduced various methods that embed hidden data into user data, e.g., by modifying the least significant bits of the digital signals – voice samples (audio) or pixels (images). Van Horenbeeck [14] proposed a steganographic method that inserts hidden data into the application layer HTTP's headers and tags.

Network steganography can use more than one protocol, in particular protocols from more than one OSI RM layer. Jankowski et al. [3] were the first to develop such a technique, called PadSteg (Padding Steganography). The term *inter-protocol steganography* has been proposed for this class of methods. PadSteg utilises the ARP, TCP, UDP or ICMP protocols, referred to as carrier-protocols, together with the Etherleak vulnerability (improper frame padding caused by ambiguities of the standardisation). To exchange steganograms, improper Ethernet frame padding is utilised in frames that use one of the carrier-protocols. While the secret communication occurs, hidden nodes can switch between carrier-protocols to minimise the risk of disclosure.

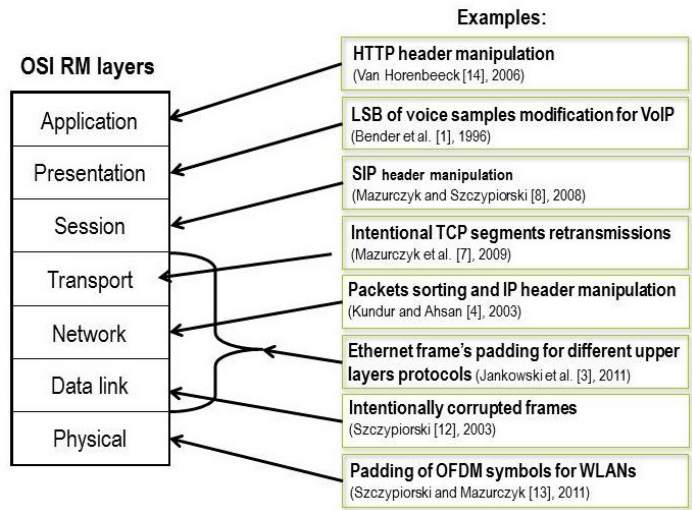

**Fig. 2** Network steganography methods classification based on protocol functions associated with OSI RM layers

Another possible classification of network steganography techniques may be based on the type of modification of Protocol Data Units (PDUs). Three types of modification may be considered: (1) modification of SDUs (Service Data Units), (2) modification of PCI (Protocol Control Information) and (3) modification of time-relations between PDUs. Case (1) includes modification of user data. Case (3) is realised, e.g., by reordering of packets and introducing purposeful delay of selected packets. Hybrid solutions involve a combination of the three cases. The classification is illustrated in Fig. 3. Hybrid solutions were introduced last; the first solutions were based on case (1), followed by solutions based on case (2).

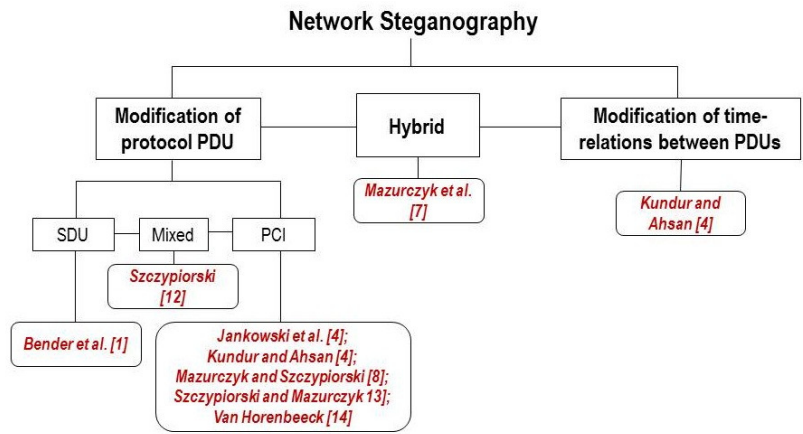

**Fig. 3** Network steganography classification based on the type of PDUs modification

In most older network steganography techniques, the place where the carrier is affected is inseparably tied to the location of the embedded steganogram. For example, in the method that is based on modification of the IP identification field [4], the modification of the carrier is tied to the location of the embedded steganogram. This tie also applies to techniques that are based on reordering of packets [4] and to techniques that modify digital media signals [1]. The tie-in was first alleviated in the HICCUPS technique [12], which resulted in increasing the technique's resistance to steganalysis.

## 4. A final remark

Due to the constantly increasing complexity of communication protocols, there is a growing need for network steganography. There is no doubt that new, more sophisticated techniques will be created and, in effect, the risk that they will be used for malicious purposes will increase. This concern adds new challenges to the difficult issue of providing network and information security. A deep understanding of the susceptibility of communication protocols to all types of manipulation (not only for steganographic purposes) becomes an extremely important issue. Research in the area of network steganography may be helpful in this respect. Research results should be used as guidelines for a methodology of designing a new generation of robust communication protocols. This is not only an engineering "must" but also an ethical obligation.